\documentclass[11pt]{article}%
\usepackage{makeidx}
\usepackage{amsfonts}
\usepackage{amsmath}
\usepackage{amssymb}
\usepackage{bm}
\usepackage{hyperref}
\usepackage{graphicx}%
\providecommand{\U}[1]{\protect\rule{.1in}{.1in}}
\numberwithin{equation}{section}

\begin{document}

\title{Zeno Effect for Bohmian Trajectories: The Unfolding of the Metatron}
\author{Maurice A. de Gosson\\\ \textit{Universit\"{a}t Wien, NuHAG}\\\textit{Fakult\"{a}t f\"{u}r Mathematik }\\\textit{A-1090 Wien}
\and Basil J. Hiley\\\ \textit{TPRU, Birkbeck}\\\textit{University of London}\\\textit{London, WC1E 7HX}}
\maketitle

\begin{abstract}
We analyse the track of an $\alpha$-particle passing through a cloud chamber
using the Bohm theory and show that the resulting classical track has its
origins in the quantum Zeno effect. By assuming the ionised gas molecules
reveal the positions of the $\alpha$-particle on its trajectory and using
these positions in a short time propagator technique developed by de Gosson,
we show it is the failure of the quantum potential to develop quickly enough
that leads to the appearance of the classical trajectory. Bohm and Hiley have
already argued that it is this failure of the quantum potential to develop
appropriately that prevents an Auger electron from undergoing a transition if
continuously watched. This allows us to conclude that, in general, it is the
suppression of the quantum potential that accounts for the quantum Zeno effect.

\end{abstract}

\section{Introduction}

Einstein writes to Bohm in 1954,

\begin{quote}
\emph{I am glad that you are deeply immersed seeking an objective description
of the phenomena and that you feel the task is much more difficult as you felt
hitherto. You should not be depressed by the enormity of the problem. If God
had created the world his primary worry was certainly not to make its
understanding easy for us. I feel it strongly since fifty years.}\cite{ae54}
\end{quote}

When David Bohm completed his book, ``Quantum Theory" \cite{dbqt}, which was
an attempt to present a clear account of Bohr's actual position, he became
dissatisfied with the overall approach \cite{db1987}. The reason for this
dissatisfaction was the fact that the theory had no place in it for an
adequate notion of an independent actuality, that is of an actual movement or
activity by which one physical state could pass over into another.

In a meeting with Einstein, ostensibly to discuss the content of his book, the
conversation eventually turned to the possibility of whether a deterministic
extension of quantum mechanics could be found. Later while exploring the WKB
approximation, Bohm realised that this approximation was giving an essentially
deterministic approach. Surely by merely truncating a series, one cannot turn
a probabilistic theory into a deterministic theory. Thus by retaining all the
terms in the series, Bohm found that one could, indeed, obtain what looked
like a deterministic description of quantum phenomena. To carry this through,
he had to assume that a quantum particle actually \emph{had} a well defined
but unknown position and momentum and followed a well-defined trajectory. This
assumption does not violate the uncertainty principle since that principle
merely states it is not possible to \emph{measure} simultaneously the position
and momentum and says nothing about whether the particle \emph{actually has} a
simultaneous position and momentum.

In the simple approach to the Bohm model, the Schr\"{o}dinger equation is
split into its real and imaginary parts with the real part showing its close
relationship to the classical Hamilton-Jacobi theory. The only difference
being the appearance of an additional term which can be regarded as a new
quality of energy, called the `quantum potential energy'. It is the properties
of this energy that enables us to account for all quantum phenomena such as,
for example, the two-slit interference effect where the trajectories are shown
to undergo a non-classical behaviour \cite{pdh}, bunching to produce the
observed fringe pattern.

Since the Schr\"{o}dinger equation still plays a defining role in the Bohm
theory, one of the key problems is to understand how the quantum potential
becomes suppressed to produce the classical world. This topic was discussed in
some detail from the global point of view in chapter 8 of Bohm and Hiley
\cite{BoHi}. In this paper we want to re-consider this problem from an
alternate point of view, focusing on an analysis of a mathematically rigorous
short time propagator technique originally developed by de Gosson
\cite{principi, Birk} to determine the quantum trajectory of a charged
particle as it passes through a gas. What we have in mind here is an $\alpha
$-particle passing through a cloud chamber, which we know leaves a track that
is essentially classical. How does this approach produce a classical
trajectory in this case?

As the $\alpha$-particle travels through the gas it leaves a trail of ions in
its wake and these ions are assumed to mark the track taken by the $\alpha
$-particle. The ions can be regarded as revealing the positions of the
$\alpha$-particle as it moves through the gas so that in a sense the particle
is being ``continuously watched" or ``monitored". If we analyse this process
from the Bohm point of view, we find that the Bohm trajectory is a classical
trajectory. Thus, in a sense, continuous observation \textquotedblleft
dequantizes\textquotedblright\ quantum trajectories. This property is, of
course, essentially an example of the quantum Zeno effect, which has been
shown to inhibit the decay of unstable quantum systems when under continuous
observation (see \cite{BoHi, fapa09,gux,Hannabuss}).

The idea lying behind the Bohm approach (Bohm and Hiley \cite{BoHi}, Hiley
\cite{Hiley1}, Hiley and collaborators \cite{hica,hicama}, Holland
\cite{Holland}) is the following: let $\Psi=\Psi(\mathbf{r},t)$ be a
wavefunction solution of Schr\"{o}dinger's equation
\[
i\hbar\frac{\partial\Psi}{\partial t}=\left[  -\frac{\hbar^{2}}{2m}%
\nabla_{\mathbf{r}}^{2}+V(\mathbf{r})\right]  \Psi.
\]
Writing $\Psi$ in polar form $\sqrt{\rho}e^{iS/\hbar}$ Schr\"{o}dinger's
equation is equivalent to the coupled systems of partial differential
equations:
\begin{equation}
\frac{\partial S}{\partial t}+\frac{(\nabla_{\mathbf{r}}S)^{2}}{2m}%
+V(\mathbf{r})+Q^{\Psi}(\mathbf{r},t)=0 \label{HJ1}%
\end{equation}
where
\begin{equation}
Q^{\Psi}=-\frac{\hbar^{2}}{2m}\frac{\nabla_{\mathbf{r}}^{2}\sqrt{|\Psi|}%
}{\sqrt{|\Psi|}}. \label{qp}%
\end{equation}
is Bohm's quantum potential (equation (\ref{HJ1}) is thus mathematically a
Hamilton-Jacobi equation), and
\begin{equation}
\frac{\partial\rho}{\partial t}+\nabla_{\mathbf{r}}\left(  \rho\frac
{\nabla_{\mathbf{r}}S}{m}\right)  =0 \label{CO1}%
\end{equation}
which is an equation of continuity that ensures the conservation of
probability. The trajectory of the particle is determined by the equation%
\begin{equation}
m\mathbf{\dot{r}}^{\Psi}=\nabla_{\mathbf{r}}S(\mathbf{r}^{\Psi},t)\text{ \ ,
\ }\mathbf{r}^{\Psi}(t_{0})=\mathbf{r}_{0} \label{Bohm}%
\end{equation}
where $\mathbf{r}_{0}$ is the initial position.

The simple derivation of equations (\ref{HJ1}) to (\ref{CO1}) obscures a
deeper mathematical relation between the Hilbert space formalism of quantum
mechanics and the Hamiltonian flows of classical mechanics. This exact
relationship has been derived in very general terms by de Gosson and Hiley
\cite{mdgbh10}, a paper that generalises the earlier work of de Gosson
\cite{Wiley, IHP, Birk}. Specifically what we show is that there is a
one-to-one and onto correspondence between Hamiltonian flows generated by a
Hamiltonian $H$ and the strongly continuous unitary one-parameter groups
satisfying Schr\"{o}dinger's equation with Hamiltonian operator $\widetilde{H}%
=H(x,-i\hbar\nabla_{x},t)$ obtained from $H$ by Weyl quantisation. This
relation exploits the \emph{metaplectic} representation of the underlying
symplectic structure \cite{Wiley,IHP,Birk}. It is the metaplectic structure
that gives rise to the quantum properties. Since the classical and quantum
motions are related but different, it was proposed in de Gosson
\cite{principi} to call the object that obeys the Bohmian law of motion
(\ref{Bohm}) a \emph{metatron}.

We choose this term rather than the usual term `particle', because we are
talking about an excitation induced by the metaplectic representation of the
underlying Hamiltonian evolution, rather than a classical object. Indeed a
deeper investigation suggests that the metatron is more like an invariant
feature of an underlying extended process, which elsewhere we have argued that
the term \emph{quantum blob} \cite{mdg04PL} may be more suggestive. However in
this paper it is sufficient to regard it as a particle-like object.

We have frequently been asked the question ``Did Bohm believe that there was
an actual classical point-like particle following these quantum trajectories?"
The answer is a definite `No'! For Bohm there was no solid `particle' either,
but instead, at the fundamental level, there was a basic process or activity
which left a `track' in, for example, the cloud chamber. Thus the track could
be explained by the \emph{enfolding} and \emph{unfolding} of an invariant form
in the overall underlying process \cite{db80}.

Thus rather than seeing the track as the continuous movement of a material
particle, it can be regarded as the continuity of a \textquotedblleft
quasi-local, semi-stable autonomous form\textquotedblright\ evolving within
this unfolding process \cite{Hiley1}. This is what we call the \emph{metatron}.

The question we will answer here is the following:

\begin{quotation}
\textit{What will the trajectory be if we continuously monitor the
metatron?}\footnote{We have deliberately chosen the word `monitored' and
avoided the word `measurement' because `measurement' in a quantum context
means, following von Neumann's Process 1 \cite{vn55}, collapsing the wave
function of the $\alpha$-particle into a position eigenfunction. Such a
process would produce a very different trajectory. In the case we are
considering here what is actually being measured in the position of the ions
and then only after the $\alpha$-particle has left the chamber. After each gas
atom is ionised, the wave function of the $\alpha$-particle takes the form
$f(\theta)e^{ik|r|}/|r|$. (See Mott \cite{nm29} and Bell \cite{jb87} for a
more detailed treatment). In our approach the information contained in this
wave function in reflected in the Hamiltonian flow $f_{t_{1} t_{2}}$ used in
Section 3.}
\end{quotation}

\section{Bohmian Trajectories Are Hamiltonian}

Let us start with the particular case where the metatron is initially
localized at a point. In this case the Bohm trajectory is Hamiltonian, a point
that we explain in section 2.2 (the general case is slightly more subtle; we
refer to the papers by Holland \cite{Holland1,Holland2} for a thorough
discussion of the interpretation of Bohmian trajectories from the Hamiltonian
point of view).

We will consider systems of $N$ material particles with the same mass $m$, and
work in generalized coordinates $\bm x=(x_{1},...,x_{n})$ and $\bm p=(p_{1}%
,...,p_{n})$, $n=3N$. Suppose that this system is sharply localized at a point
$\bm x_{0}=(x_{1,0},...,x_{n,0})$ at time $t_{0}$. The classical Hamiltonian
function is
\begin{equation}
H(\bm x,\bm p)=\frac{\bm p^{2}}{2m}+V(\bm x) \label{hamf}%
\end{equation}
hence the organising field of this system is the solution of the
Schr\"{o}dinger equation
\begin{equation}
i\hbar\frac{\partial\Psi}{\partial t}=\left[  -\frac{\hbar^{2}}{2m}%
\nabla_{\bm x}^{2}+V(\bm x)\right]  \Psi\text{ \ , \ }\Psi(\bm x,t_{0}%
)=\delta(\bm x-\bm x_{0}) \label{schr}%
\end{equation}
where $\nabla_{\bm x}$ is the $n$-dimensional gradient in the variables
$x_{1},...,x_{n}$. The function $\Psi$ is thus just the propagator
$G(\bm x,\bm x_{0};t,t_{0})$ of the Schr\"{o}dinger equation. We write $G$ in
polar form%
\[
G(\bm x,\bm x_{0};t,t_{0})=\sqrt{\rho(\bm x,\bm x_{0};t,t_{0})}e^{\frac
{i}{\hbar}S(\bm x,\bm x_{0};t,t_{0})}.
\]
The equation of motion (\ref{Bohm}) is in this case
\begin{equation}
m\dot{\bm x}^{\Psi}=\nabla_{\bm x}S(\bm x^{\Psi},\bm x_{0};t,t_{0})\text{ \ ,
\ }\bm x^{\Psi}(t_{0})=\bm x_{0}. \label{xo}%
\end{equation}

\subsection{Short-time estimates}

We are going to give a short-time estimate for the function $S$. The interest
of this estimate is two-fold: it will not only allow us to give a precise
statement of the Zeno effect for Bohmian trajectories, but it will also allow
us to prove in detail the Hamiltonian character of these trajectories.

We will assume that the potential $V$ is at least twice continuously
differentiable in the variables $x_{1},...,x_{n}$.

In \cite{principi}, Chapter 7, we established the following short-time
formulas for $t-t_{0}\rightarrow0$ (a similar formula has been obtained in
\cite{makmil1,makmil2,Schiller1,Schiller2}):
\begin{equation}
S(\bm x,\bm x_{0};t,t_{0})=\sum_{j=1}^{n}\frac{m(x_{j}-x_{0,j})^{2}}%
{2(t-t_{0})}-\tilde{V}(\bm x,\bm x_{0})(t-t_{0})+O((t-t_{0})^{2}) \label{wapp}%
\end{equation}
where $\tilde{V}(\bm x,\bm x^{\prime})$ is the average value of the potential
on the line segment $[\bm x^{\prime},\bm x]$:%
\begin{equation}
\tilde{V}(\bm x,\bm x_{0})=\int_{0}^{1}V(\lambda\bm x+(1-\lambda
)\bm x_{0})d\lambda. \label{utilde}%
\end{equation}

We observe that the quantum potential is absent from formula (\ref{wapp}); we
would actually have obtained the same approximation if we had replaced $S$
with the solution to the classical Hamilton--Jacobi equation
\[
\frac{\partial S_{\text{cl}}}{\partial t}+\frac{(\nabla_{\bm x}S)^{2}}%
{2m}+V(\bm x)=0
\]
while $S$ is actually a solution of the quantum Hamilton--Jacobi equation
\begin{equation}
\frac{\partial S}{\partial t}+\frac{(\nabla_{\bm x}S)^{2}}{2m}%
+V(\bm x)+Q^{\Psi}(\bm x,t)=0.
\end{equation}
\ How can this be? The reason is that if we replace the propagator
$G(\bm x,\bm x_{0};t,t_{0})$ by its \textquotedblleft
classical\textquotedblright\ approximation
\[
G_{\text{cl}}(\bm x,\bm x_{0};t,t_{0})=\sqrt{\rho_{\text{cl}}(\bm x,\bm x_{0}%
;t,t_{0})}e^{\frac{i}{\hbar}S_{\text{cl}}(\bm x,\bm x_{0};t,t_{0})}%
\]
where $\rho_{\text{cl}}$ is the Van Vleck density (i.e. the determinant of the
matrix of second derivatives of $S_{\text{cl}}$) then we have
\[
G(\bm x,\bm x_{0};t,t_{0})-G_{\text{cl}}(\bm x,\bm x_{0};t,t_{0}%
)=O((t-t_{0})^{2})
\]
(cf. Lemma 241 in \cite{principi}) from which follows that%
\[
-\frac{\hbar^{2}}{2m}\frac{\nabla_{\bm x}^{2}G}{G}-\left(  -\frac{\hbar^{2}%
}{2m}\right)  \frac{\nabla_{\bm x}^{2}G_{\text{cl}}}{G_{\text{cl}}}%
=O((t-t_{0})^{2});
\]
the difference between these two terms, $O((t-t_{0})^{2})$, is thus absorbed
by the corresponding term in (\ref{wapp}). [We take the opportunity to remark
that when the potential $V(\bm x)$ is quadratic in the position variables
$x_{1},...,x_{n}$ then $G_{\text{cl}}=G$; we will come back to this relation
later in section 3.1].

Moreover, formula (\ref{wapp}) can be twice continuously differentiated with
respect to the variables $\bm x_{j}$ and $\bm x_{0,j}$. It follows that the
second derivatives of $S$ are given by%
\[
\frac{\partial^{2}S}{\partial x_{j}\partial x_{0,k}}=\frac{m}{t-t_{0}}%
\delta_{jk}+O(t-t_{0})
\]
and hence the Hessian matrix $S_{x,x_{0}}$ (i.e. the matrix of mixed second
derivatives) satisfies
\begin{equation}
\det(S_{\bm x,\bm x_{0}})=\left(  \frac{m}{t-t_{0}}\right)  ^{n}+O(t-t_{0}).
\label{hessian}%
\end{equation}

Formula (\ref{wapp}) is the key to the following important asymptotic version
of Bohm's equation (\ref{xo}):
\begin{equation}
\dot{\bm x}^{\Psi}=\frac{\bm x^{\Psi}-\bm x_{0}}{t-t_{0}}-\frac{1}{2m}%
\nabla_{\bm x}V(\bm x_{0})(t-t_{0})+O((t-t_{0})^{2})). \label{master}%
\end{equation}

Let us prove this formula. Using the expansion (\ref{wapp}), formula
(\ref{xo}) becomes%
\begin{equation}
\dot{\bm x}^{\Psi}=\frac{\bm x^{\Psi}-\bm x_{0}}{t-t_{0}}-\frac{1}{m}%
\nabla_{\bm x}\tilde{V}(\bm x^{\Psi},\bm x_{0})(t-t_{0})+O((t-t_{0})^{2}).
\label{eqn}%
\end{equation}
Let us show that
\begin{equation}
\nabla_{\bm x}\tilde{V}(\bm x^{\Psi},\bm x_{0})=\frac{1}{2}\nabla
_{\bm x}V(\bm x_{0})+O(t-t_{0}); \label{nablax}%
\end{equation}
this will complete the proof of formula (\ref{master}). We first note that
(\ref{eqn}) implies in particular that
\[
\dot{\bm x}^{\Psi}=\frac{\bm x^{\Psi}-\bm x_{0}}{t-t_{0}}+O(t-t_{0})
\]
and thus $\bm x^{\Psi}$ is given by%
\begin{equation}
\bm x^{\Psi}(t)=\bm x_{0}+\frac{\bm p_{0}}{m}(t-t_{0})+O((t-t_{0})^{2})
\label{xpsi1}%
\end{equation}
where $p_{0}$ is an arbitrary constant vector. In particular we have
$O(\bm x^{\Psi}-\bm x_{0})=O(t-t_{0})$ and hence
\begin{align*}
\nabla_{\bm x}\tilde{V}(\bm x^{\Psi},\bm x_{0})  &  =\nabla_{\bm x}\tilde
{V}(\bm x_{0},\bm x_{0})+O(\bm x^{\Psi}-\bm x_{0})\\
&  =\nabla_{\bm x}\tilde{V}(\bm x_{0},\bm x_{0})+O(t-t_{0})
\end{align*}
from which it follows that
\begin{align*}
\nabla_{\bm x}\tilde{V}(\bm x^{\Psi},\bm x_{0})  &  =\int_{0}^{1}\lambda
\nabla_{\bm x}V(\lambda\bm x_{0}+(1-\lambda)\bm x_{0})d\lambda+O(t-t_{0})\\
&  =\frac{1}{2}\nabla_{\bm x}V(\bm x_{0})+O(t-t_{0})
\end{align*}
which is precisely the estimate (\ref{nablax}).

\subsection{The Hamiltonian character of Bohmian trajectories}

Let $\bm p_{0}=(p_{1,0},...,p_{n,0})$ be an arbitrary momentum vector, and set%
\begin{equation}
\bm p_{0}=-\nabla_{\bm x_{0}}S(\bm x,\bm x_{0};t,t_{0}). \label{po}%
\end{equation}
In view of formula (\ref{hessian}), the Hessian of $S$ in the variables $x$
and $x_{0}$ is invertible for small values of $t$, hence the implicit function
theorem implies that (\ref{po}) determines a function $\bm x=\bm x(t)$
(depending on $\bm x_{0}$ and $t_{0}$ viewed as parameters), defined by
\begin{equation}
\bm p_{0}=-\nabla_{\bm x_{0}}S(\bm x(t),\bm x_{0};t,t_{0}). \label{pot}%
\end{equation}
Setting
\begin{equation}
\bm p(t)=\nabla_{\bm x}S(\bm x(t),\bm x_{0};t,t_{0}) \label{pt}%
\end{equation}
we claim that the functions $x(t)$ and $p(t)$ thus defined are solutions of
the Hamilton equations
\begin{equation}
\dot{\bm x}=\nabla_{\bm p}H^{\Psi}(\bm x,\bm p,t)\text{ \ , \ }\dot
{\bm p}=-\nabla_{\bm x}H^{\Psi}(\bm x,\bm p,t) \label{Hamilton}%
\end{equation}
and that we have $\bm x(t_{0})=\bm x_{0}$, $\bm p(t_{0})=\bm p_{0}$. We are
actually going to use classical Hamilton--Jacobi theory (see
\cite{Arnold,HGoldstein,principi,Birk} or any introductory text on analytical
mechanics). For notational simplicity we assume that $n=1$. The function $S$
satisfies the equation%
\begin{equation}
\frac{\partial S}{\partial t}+\frac{1}{2m}\left(  \frac{\partial S}{\partial
x}\right)  ^{2}+V(x)-\frac{\hbar^{2}}{2m}\frac{1}{\sqrt{\rho}}\frac
{\partial^{2}\sqrt{\rho}}{\partial x^{2}}=0; \label{HJ2}%
\end{equation}
introducing the quantum potential%
\begin{equation}
Q^{\Psi}=-\frac{\hbar^{2}}{2m}\frac{1}{\sqrt{\rho}}\frac{\partial^{2}%
\sqrt{\rho}}{\partial x^{2}} \label{qpn}%
\end{equation}
we set $H^{\Psi}=H+Q^{\Psi}$ so that (\ref{HJ2}) is just the quantum
Hamilton--Jacobi equation%
\begin{equation}
\frac{\partial S}{\partial t}+H^{\Psi}\left(  x,\frac{\partial S}{\partial
x},t\right)  =0. \label{HJ3}%
\end{equation}
Differentiating the latter with respect to $p=\partial S/\partial x$ yields,
using the chain rule,%
\begin{equation}
\frac{\partial^{2}S}{\partial x_{0}\partial t}+\frac{\partial H^{\Psi}%
}{\partial p}\frac{\partial^{2}S}{\partial x_{0}\partial x}=0 \label{a}%
\end{equation}
and differentiating the equation (\ref{pot}) with respect to time yields%
\begin{equation}
\frac{\partial^{2}S}{\partial x_{0}\partial t}+\frac{\partial^{2}S}{\partial
x\partial x_{0}}\dot{x}=0. \label{b}%
\end{equation}
Subtracting (\ref{b}) from (\ref{a}) we get%
\[
\frac{\partial^{2}S}{\partial x\partial x_{0}}\left(  \frac{\partial H^{\Psi}%
}{\partial p}-\dot{x}\right)  =0
\]
which produces the first Hamilton equation (\ref{Hamilton}) since it is
assumed that we have $\partial^{2}S/\partial x\partial x_{0}\neq0$. Let us
next show that the second Hamilton equation (\ref{Hamilton}) is satisfied as
well. For this we differentiate the quantum Hamilton--Jacobi equation
(\ref{HJ3}) with respect to $x$, which yields%
\begin{equation}
\frac{\partial^{2}S}{\partial x\partial t}+\frac{\partial H^{\Psi}}{\partial
x}+\frac{\partial H^{\Psi}}{\partial p}\frac{\partial^{2}S}{\partial x^{2}}=0.
\label{c}%
\end{equation}
Differentiating the equality (\ref{pt}) with respect to $t$ we get%
\begin{equation}
\frac{\partial^{2}S}{\partial t\partial x}=-\dot{p}(t)-\frac{\partial^{2}%
S}{\partial x^{2}}\dot{x} \label{d}%
\end{equation}
and hence the equation (\ref{c}) can be rewritten%
\[
-\dot{p}(t)-\frac{\partial^{2}S}{\partial x^{2}}\dot{x}+\frac{\partial
H^{\Psi}}{\partial x}+\frac{\partial H^{\Psi}}{\partial p}\frac{\partial^{2}%
S}{\partial x^{2}}=0.
\]
Taking into account the relation $\dot{x}=\partial H^{\Psi}/\partial p$
established above we have
\[
-\dot{p}(t)-\frac{\partial H^{\Psi}}{\partial x}=0
\]
which is precisely the second Hamilton equation (\ref{Hamilton}). There
remains to show that we have $x(t_{0})=x_{0}$ and $p(t_{0})=p_{0}$. Recall
that $x(t)$ is defined by the implicit equation
\[
p_{0}=-\nabla_{x_{0}}S(x(t),x_{0};t,t_{0})
\]
(equation (\ref{pot})); in view of the short-time estimate (\ref{wapp}) this
means that we have%
\[
p_{0}=\frac{m(x(t)-x_{0})}{t-t_{0}}+O(t-t_{0})
\]
and hence we must have $\lim_{t\rightarrow t_{0}}x(t)=x(t_{0})=x_{0}$. This
also implies that $p_{0}=m\dot{x}(t_{0})=p(t_{0})$.

In conclusion we have thus shown that:

\begin{quote}
\emph{Bohm's equation of motion (\ref{xo}) is equivalent to Hamilton's
equations (\ref{Hamilton}).}
\end{quote}

To complete our discussion, we make two important observations:

\begin{itemize}
\item Even when the Hamiltonian function $H$ does not depend explicitly on
time, the function $H^{\Psi}=H+Q^{\Psi}$ is usually time-dependent (because
the quantum potential generally is), so the flow $(f_{t}^{\Psi})$ it
determines does not inherit the usual group property $f_{t}f_{t^{\prime}%
}=f_{t+t^{\prime}}$ of the flow determined by the classical Hamiltonian $H$.
One has instead to use the \textquotedblleft time-dependent
flow\textquotedblright\ $(f_{t,t^{\prime}}^{\Psi})$, which has a groupoid
property in the sense that $f_{t,t^{\prime}}^{\Psi}f_{t^{\prime}%
,t^{\prime\prime}}^{\Psi}=f_{t,t^{\prime\prime}}^{\Psi}$.

\item The time-dependent flow\ $(f_{t,t^{\prime}}^{\Psi})$ consists of
canonical transformations; that is, the Jacobian matrix of $f_{t,t^{\prime}%
}^{\Psi}$ calculated at any point $(x,p)$ where it is defined by a symplectic
matrix. This is an immediate consequence of the fact discussed above, namely,
that the flow determined by \emph{any} Hamiltonian function has this property.
\end{itemize}

We have seen that the Bohmian trajectory for a particle initially sharply
localized at a point $\bm x_{0}$ is Hamiltonian, and in fact governed by the
Hamilton equations (\ref{Hamilton}):%
\begin{equation}
\dot{\bm x}=\nabla_{\bm p}H^{\Psi}(\bm x,\bm p,t)\text{ \ , \ }\dot
{\bm p}=-\nabla_{\bm x}H^{\Psi}(\bm x,\bm p,t). \label{hameq}%
\end{equation}
The discussion of short-time solutions of Bohm's equation of motion allows us
to give approximations to the solution. First, the solutions of the equation
$\dot{\bm x}=\nabla_{\bm p}H^{\Psi}(\bm x,\bm p,t)$ are given by the simple
relation%
\[
\bm x^{\Psi}(t)=\bm x_{0}+\frac{\bm p_{0}}{m}(t-t_{0})+O((t-t_{0})^{2})
\]
as was already noticed in (\ref{xpsi1}). Then we proved that the momentum
$\bm p^{\Psi}(t)=m\dot{\bm x}^{\Psi}(t)$ is given by equation (\ref{master}):%
\begin{equation}
m\dot{\bm x}^{\Psi}(t)=\frac{m(\bm x^{\Psi}(t)-\bm x_{0})}{t-t_{0}}-\frac
{1}{2}\nabla_{\bm x}V(\bm x_{0})(t-t_{0})+O((t-t_{0})^{2}). \label{pdot}%
\end{equation}
However we cannot solve this equation by inserting the value of $x^{\Psi}(t)$
above since this would lead to an estimate modulo $O(t-t_{0})$ not
$O((t-t_{0})^{2})$. What we do is the following: differentiating both sides of
the equation (\ref{pdot}) with respect to $t$ we get%
\[
\ddot{\bm x}^{\Psi}(t)=\frac{\bm x^{\Psi}(t)-\bm x_{0}}{(t-t_{0})^{2}}%
+\frac{\dot{\bm x}^{\Psi}(t)}{t-t_{0}}-\frac{1}{2m}\nabla_{\bm x}%
V(\bm x_{0})+O(t-t_{0})
\]
that is, replacing $\dot{\bm x}^{\Psi}(t)$ by the value given by (\ref{pdot}),%
\[
\dot{\bm p}^{\Psi}(t)=m\ddot{\bm x}^{\Psi}(t)=-\nabla_{\bm x}V(\bm  x_{0}%
)+O(t-t_{0}).
\]
Solving this equation we get
\[
\bm p^{\Psi}(t)=\bm p_{0}-\nabla_{\bm x}V(\bm x_{0})(t-t_{0})+O((t-t_{0}%
)^{2}).
\]
Summarizing, the solutions of the Hamilton equations (\ref{hameq}) for
$H^{\Psi}=H+Q^{\Psi}$ are given by%
\begin{align}
\bm x^{\Psi}(t)  &  =\bm x_{0}+\frac{\bm p_{0}}{m}(t-t_{0})+O((t-t_{0}%
)^{2})\label{eq:29}\\
\bm p^{\Psi}(t)  &  =\bm p_{0}-\nabla_{\bm x}V(\bm x_{0})(t-t_{0}%
)+O((t-t_{0})^{2}).\label{eq:30}%
\end{align}
The observant reader will have noticed that (up to the error term
$O((t-t_{0})^{2})$) there is no trace of the quantum potential $Q^{\Psi}$ in
these short-time formulas. Had we replaced the function $H^{\Psi}$ with the
classical Hamiltonian $H$ we would actually have obtained exactly the same
solutions, up to the $O((t-t_{0})^{2})$ term.

\section{Bohmian Zeno Effect}

\subsection{The case of quadratic potentials}

Here is an easy case; it is in fact so easy that it is slightly misleading:
the Bohmian trajectories are here classical trajectories from the beginning,
because the quantum potential vanishes.

Let us assume that the potential $V(\bm x)$ is a quadratic form in the
position variables, that is
\[
V(\bm x)=\frac{1}{2}M\bm x\cdot\bm x
\]
where $M$ is a symmetric matrix. Using the theory of the metaplectic
representation \cite{Wiley,IHP,principi,Birk} it is well-known that the
propagator $G$ is given by the formula
\begin{equation}
G(\bm x,\bm x_{0};t,t_{0})=\left(  \tfrac{1}{2\pi i\hbar}\right)
^{n/2}i^{m(t,t_{0})}\sqrt{|\rho(t,t_{0})|}e^{\frac{i}{\hbar}W(\bm x,\bm x_{0}%
;t,t_{0})} \label{Green}%
\end{equation}
where $W(\bm x,\bm x_{0};t,t_{0})$ is Hamilton's two-point characteristic
function (see e.g. \cite{Arnold,HGoldstein}): it is a quadratic form%
\[
W=\frac{1}{2}P\bm x\cdot\bm x-L\bm x\cdot\bm  x_{0}+\frac{1}{2}B\bm x_{0}%
\cdot\bm x_{0}%
\]
where $P=P(t,t_{0})$ and $B=B(t,t_{0})$ are symmetric matrices and
$L=L(t,t_{0})$ is invertible; viewed as function of $x$ it satisfies the
Hamilton--Jacobi equation%
\[
\frac{\partial W}{\partial t}+\frac{(\nabla_{\bm x}W)^{2}}{2m}+\frac{1}%
{2}M\bm x\cdot\bm x.
\]
Moreover, $m(t,t_{0})$ in equation (\ref{Green}) is an integer
(\textquotedblleft Maslov index\textquotedblright) and $\rho(t,t_{0})$ is the
determinant of $L=L(t,t_{0})$ (the Van Vleck density). Since $m(t,t_{0})$ and
$\rho(t,t_{0})$ do not depend on $x$, it follows that the quantum potential
$Q^{\Psi}$ determined by the propagator (\ref{Green}) is zero. Since we have
$H^{\Psi}=H+Q^{\Psi}$, we see immediately that the quantum motion is perfectly
classical in this case: the quantum equations of motion (\ref{Hamilton})
reduce to the ordinary Hamilton equations%
\begin{equation}
\dot{\bm x}=\frac{\bm p}{m}\ ,\ \dot{\bm p}=-M\bm x \label{qua}%
\end{equation}
which can be easily integrated: in particular the flow $(f_{t})$ they
determine is a true flow (because $H=H^{\Psi}$ is time-independent) and
consists of symplectic matrices (\cite{Arnold,principi,Birk,HGoldstein}). In
fact,%
\[
f_{t}=e^{tX}\text{ \ , \ }X=%
\begin{pmatrix}
0_{n\times n} & \frac{1}{m}I_{n\times n}\\
-M & 0_{n\times n}%
\end{pmatrix}
.
\]
Thus, in the case of quadratic potentials the Bohmian trajectories associated
with the propagator are the usual Hamilton trajectories associated with the
classical Hamiltonian function of the problem.

Suppose now that we monitor \textquotedblleft continuously\textquotedblright%
(in the sense discussed in the introduction) the time evolution of the
metatron --which is so far \textquotedblleft quantum\textquotedblright-- and
try to find out what effect this interaction has on the trajectory. In the
example of the cloud chamber, let $\Delta t$ be the time between successive
ionisations. From the mathematical point of view we will assume the limit
$\Delta t\rightarrow0$ exists and that it is continuous and smooth. In other
words we are neglecting the reaction of the ion formation on the $\alpha
$-particle, an assumption that Mott \cite{nm29} also makes. Thus we can assign
at every point a velocity vector.

Let us choose a time interval $[0,t]$ (typically $t=1%
\operatorname{s}%
$) and subdivide it in a sequence of $N$ intervals
\begin{align}
\lbrack0,\Delta t]\text{ }[\Delta t,2\Delta t]\text{ }[2\Delta t,3\Delta
t]\cdot\cdot\cdot\lbrack(N-1)\Delta t,N\Delta t]\label{eq:SEQ}%
\end{align}
with $\Delta t=t/N$; the integer $N$ is assumed to be very large (for instance
$N \simeq10^{6}-10^{8}$). Assume that at time $t_{0}=0$ the particle is at a
point $\bm x_{0}$ and after after time $\Delta t$ it is at $\bm x_{1}$ ; its
momentum is $\bm p_{1}$ and we have $(\bm x_{1},\bm p_{1})=f_{\Delta
t}(\bm x_{0},\bm p_{0})$. We now repeat the procedure, replacing $\bm x_{0}$
by $\bm x_{1}$; since the trajectory is assumed to be smooth, the initial
momentum will be $\bm p_{1}$ and after time $\Delta t$ the particle will be at
$\bm x_{2}$ with momentum $\bm p_{2}$ such that $(\bm x_{2},\bm p_{2}%
)=f_{\Delta t}(\bm x_{1},\bm p_{1})=f_{\Delta t}f_{\Delta t}(\bm x_{0}%
,\bm p_{0})$. Repeating the same process until time $t=N\Delta t$ we find a
series of points in space which the particle takes as positions one after
another\footnote{In conformity with W. Heisenberg's statement:
\textquotedblleft By path we understand a series of points in space which the
electron takes as `positions' one after another\textquotedblright%
\ \cite{wh27}} that $(\bm x_{N},\bm p_{N})=(f_{\Delta t})^{N}(\bm x_{0}%
,\bm p_{0})$. But in view of the group property $f_{t}f_{t^{\prime}%
}=f_{t+t^{\prime}}$ of the flow we have $(f_{\Delta t})^{N}=f_{N\Delta
t}=f_{t}$ and hence $(\bm x_{N},\bm p_{N})=f_{t}(\bm x_{0},\bm p_{0})$. The
observed Bohmian trajectory is thus the classical trajectory predicted by
Hamilton's equations.

\subsection{The general case}

In generalizing the discussion above to arbitrary potentials, $V(\bm x)$,
there are two difficulties. The first is that we do not have exact equations
for the Bohmian trajectory, but only short-time approximations. The second is
that the Hamilton equations for $\bm x^{\Psi}$ and $\bm p^{\Psi}$ no longer
determine a flow having a group property because the Hamiltonian $H^{\Psi}$ is
time-dependent. Nevertheless the material we have developed so far is actually
sufficient to show that the observed trajectory is the classical one.

The key will be the theory of Lie--Trotter algorithms which is a powerful
method for constructing exact solutions from short-time estimates. The method
goes back to early work of Trotter \cite{Trotter} elaborating on Sophus Lie's
proof of the exponential matrix formula $e^{A+B}=\lim_{N\rightarrow\infty
}\left(  e^{A/N}e^{B/N}\right)  ^{N}$; see Chorin et al. \cite{chorinetal} for
a detailed and rigorous study; we have summarized the main ideas in the
Appendix B of \cite{principi}); also see Nelson \cite{Nelson}. (We mention
that there exists an operator variant of this procedure, called the
Trotter--Kato formula.)

Let us begin by introducing some notation. We have seen that the datum of the
propagator $G_{0}=G(\bm x,\bm x_{0};t,t_{0})$ determines a quantum potential
$Q^{\Psi}$ and thus Hamilton equations (\ref{Hamilton}) associated with
$H^{\Psi}=H+Q^{\Psi}$. We now choose $t_{0}=0$ and denote the corresponding
quantum potential by $Q^{0}$ and set $H^{0}=H+Q^{0}$. After time $\Delta t$
the position of the particle is at $\bm x_{1}$. The future evolution of the
particle is now governed by the new propagator $G_{1}=G(\bm x,\bm x_{1}%
;t,t_{0})$, leading to a new quantum potential $Q^{1}$ and to a new
Hamiltonian $H^{1}$; repeating this until time $t$ we thus have a sequence of
points $\bm x_{0},\bm x_{1},...,\bm x_{N}=\bm x$ and a corresponding sequence
of Hamiltonian functions $H^{0},H^{1},...,H^{N}$ determined by the quantum
potentials $Q^{0},Q^{1},...,Q^{N}$. We denote by $(f_{t,t_{0}}^{0})$,
$(f_{t,t_{1}}^{1})$,...,$(f_{t,t_{N-1}}^{N-1})$ the time dependent flows
determined by the Hamiltonian functions $H^{0},H^{1},...,H^{N}$; we have set
here $t_{1}=t_{0}+\Delta t$, $t_{2}=t_{1}+\Delta t$ and so on.

Repeating the procedure explained in the case of quadratic potentials, we get
in this case a sequence of successive equalities%
\begin{gather*}
(\bm x_{1},\bm p_{1})=f_{t_{1},t_{0}}^{0}(\bm x_{0},\bm p_{0})\\
(\bm x_{2},\bm p_{2})=f_{t_{2},t_{1}}^{1}(\bm x_{1},\bm p_{1})\\
\cdot\cdot\cdot\cdot\cdot\cdot\cdot\cdot\cdot\cdot\\
(\bm x,\bm p)=f_{t,t_{N-1}}^{N-1}(\bm x_{N-1},\bm p_{N-1})
\end{gather*}
which implies that the final position $\bm x=\bm x_{N}$ at time $t$ is
expressed in terms of the initial point $\bm x_{0}$ by the formula%
\[
(\bm x,\bm p)=f_{t,t_{N-1}}^{N-1}\cdot\cdot\cdot f_{t_{2},t_{1}}^{1}%
f_{t_{1},t_{0}}^{0}(\bm x_{0},\bm p_{0}).
\]
Denote now by $(g_{t,t_{0}}^{0})$, $(g_{t,t_{1}}^{1})$,...,$(g_{t,t_{N-1}%
}^{N-1})$ the approximate flows determined by the equations%
\begin{align*}
(\bm x_{1},\bm p_{1})  &  =(\bm x_{0}+\frac{\bm p_{0}}{m}\Delta t,\bm p_{0}%
-\nabla_{\bm x}V(\bm x_{0})\Delta t)\\
(\bm x_{2},\bm p_{2})  &  =(\bm x_{1}+\frac{\bm p_{1}}{m}\Delta t,\bm p_{1}%
-\nabla_{x}V(\bm x_{1})\Delta t)\\
&  \cdot\cdot\cdot\cdot\cdot\cdot\cdot\cdot\cdot\cdot\\
(\bm x,\bm p)  &  =(\bm x_{N-1}+\frac{\bm p_{N-1}}{m}\Delta t,\bm p_{N-1}%
-\nabla_{\bm x}V(\bm x_{N-1})\Delta t).
\end{align*}
Invoking the Lie--Trotter formula, the sequence of estimates
\[
f_{t_{k},t_{k-1}}^{0}(\bm x_{k-1},\bm p_{k-1})-g_{t_{k},t_{k-1}}%
^{0}(\bm x_{k-1},\bm p_{k-1})=O(\Delta t^{2})
\]
implies that we have%
\[
\lim_{N\rightarrow\infty}g_{t,t_{N-1}}^{N-1}\cdot\cdot\cdot g_{t_{2},t_{1}%
}^{1}g_{t_{1},0}^{0}(\bm x_{0},\bm p_{0})=\lim_{N\rightarrow\infty
}f_{t,t_{N-1}}^{N-1}\cdot\cdot\cdot f_{t_{2},t_{1}}^{1}f_{t_{1},0}%
^{0}(\bm x_{0},\bm p_{0})
\]
The argument goes as follows (for a detailed proof see \cite{principi}): since
we have $g_{t_{k},t_{k-1}}^{k}=f_{t_{k},t_{k-1}}^{k}+O(\Delta t^{2})$ the
product is approximated by%
\[
g_{t,t_{N-1}}^{N-1}\cdot\cdot\cdot g_{t_{2},t_{1}}^{1}g_{t_{1},t_{0}}%
^{0}=f_{t,t_{N-1}}^{N-1}\cdot\cdot\cdot f_{t_{2},t_{1}}^{1}f_{t_{1},t_{0}}%
^{0}+NO(\Delta t^{2})
\]
and since $\Delta t=t/N$ we have $NO(\Delta t^{2})=O(\Delta t)$ which goes to
zero when $N\rightarrow\infty$.

Now, recall our remark that the quantum potential is absent from the
approximate flows $g_{t_{k},t_{k-1}}^{k}$; using again the Lie--Trotter
formula together with short-time approximations to the Hamiltonian flow
$(f_{t})$ determined by the classical Hamiltonian $H$, we get
\[
\lim_{N\rightarrow\infty}g_{t,t_{N-1}}^{N-1}\cdot\cdot\cdot g_{t_{2},t_{1}%
}^{1}g_{t_{1},0}^{0}(\bm x_{0},\bm p_{0})=f_{t}%
\]
and hence%
\[
\lim_{N\rightarrow\infty}f_{t,t_{N-1}}^{N-1}\cdot\cdot\cdot f_{t_{2},t_{1}%
}^{1}f_{t_{1},0}^{0}(\bm x_{0},\bm p_{0})=f_{t}%
\]
which shows that the trajectory is the classical one.

\section{Conclusion.}

In this paper we have shown how a detailed mathematical examination of the
deeper symplectic structure that underlies the Bohm approach predicts that if
a quantum particle is monitored continuously in the way we have suggested, it
will follow a classical trajectory.

The idea lying behind this result becomes clear once one realises that it is
the appearance of the quantum potential energy that distinguishes quantum
behaviour from classical behaviour. Indeed this is very obvious if we examine
equation (\ref{HJ1}) and compare it with the classical Hamilton-Jacobi
equation. The essential difference is the appearance of the term, $Q^{\Psi}$,
in equation (\ref{HJ1}). This means that when $Q^{\Psi}$ is negligible
compared with the kinetic energy, the equation simply reduces to the classical
Hamilton-Jacobi equation.

In this paper we have shown that the suppression of the quantum potential is
possible if the successive positions of the particle can be defined in a short
enough time. To see this, we must examine equations (\ref{eq:29}) and
(\ref{eq:30}), which are exact to $O\left( (\Delta t)^{2}\right) $. Notice
that there is no quantum potential present in either equation. Only when we
allow higher order terms does the quantum potential appear. Thus if it is
possible to obtain information of succession of positions in a short enough
time without deflecting the particle significantly, then equation
(\ref{eq:SEQ}) shows that no quantum potential will appear and the trajectory
will be a classical trajectory. In other words the quantum Zeno effect arises
because $Q^{\Psi}$ is prevented from contributing to the process.

Another illustration of how continuous observations of a different kind can
give rise to a quantum Zeno effect has already been given in Bohm and Hiley
\cite{BoHi}. They considered the transition of an Auger-like particle and
showed that the perturbed wave function, which is proportional to $\Delta t$
for times less that $1/\Delta E$, ($\Delta E$ is the energy released in the
transition) will never become large and therefore cannot make a significant
contribution to the quantum potential necessary for the transition to occur.
Thus again for the reason that no transition will take place is the vanishing
of the quantum potential.

Our discussion shows that the Bohm model has a very different way of arriving
at the classical limit than the prevailing view based on decoherence. In our
view the main difficulty in using decoherence is that it merely destroys the
off-diagonal elements of the density matrix but it does not explain how the
classical equations of motion arise. It continues to describe classical
objects using wave functions, a criticism that has already been made by Primas
\cite{primas}.

The mathematics we have used in this paper is a further example of how the
relation between the symplectic and metaplectic representations discussed in
our earlier paper \cite{mdgbh10} holds a further clue of the relationship
between the quantum and classical domains.  It is when the global properties
of the covering (metaplectic) group become unimportant that the classical
world emerges. As has been pointed out by Hiley \cite{bh03} \cite{bh09}, the
Bohm approach has a close relationship to the Moyal approach. This supplements
the work of de Gosson \cite{Birk} who shows exactly how the Wigner-Moyal
transformation is related to the mathematical structure we are exploiting
here. The Moyal approach involves a deformed Poisson algebra from which the
classical limit emerges in a very simple way, namely, in those situations
where the deformation parameter can be considered to be small which is
essentially similar to neglecting the quantum potential.


\begin{thebibliography}{99}                                                                                               %


\bibitem {AM}Abraham, R. and Marsden, J.E.: \emph{Foundations of
Mechanics.}\textit{\ }The Benjamin/Cummings Publishing Company, 2nd edition, (1978)

\bibitem {Arnold}Arnold, V.I.: Mathematical Methods of Classical Mechanics.
\emph{Graduate Texts in Mathematics}, second edition, Springer-Verlag, (1989)

\bibitem {jb87}Bell, J. S. \emph{Speakable and Unspeakable in Quantum
Mechanics,} Cambridge University Press, Cambridge, 1987.

\bibitem {dbqt}Bohm, D., \emph{Quantum Theory} Prentice-Hall, Englewood
Cliffs, N. J., 1951.

\bibitem {Bohm}Bohm, D.: A Suggested Interpretation of the Quantum Theory in
Terms of Hidden Variables. \emph{Phys. Rev.} \textbf{85} (1952), 166--179, 180--193

\bibitem {db1987}Bohm, D., Hidden Variables and the Implicate Order, in Hiley,
B. J. and Peat, D. F., \emph{Quantum Implications: Essays in Honour of David
Bohm}, Routledge, London, 1987.

\bibitem {db80}Bohm, D., \emph{Wholeness and the Implicate Order}, Routledge,
London, 1980.

\bibitem {dbbh85}Bohm, D. and Hiley, B. J., Unbroken Quantum Realism, from
Microscopic to Macroscopic Levels. \emph{Phys. Rev. Letters}, \textbf{55},
(1985), 2511-1514.

\bibitem {BoHi}Bohm, D., Hiley, B.: \emph{The Undivided Universe: An
Ontological Interpretation of Quantum Theory}. London \& New York: Routledge (1993)

\bibitem {bohi87}Bohm, D., Hiley, B.: An Ontological Basis for the Quantum
Theory: I-Nonrelativistic Particle Systems. \emph{Phys. Reports} \textbf{144}
(1987), 323--348

\bibitem {chorinetal}Chorin, A.J., Hughes T.J.R, McCracken, M.F., Marsden,
J.E.: Product formulas and numerical algorithms. \emph{Comm. Pure and Appl.
Math.} \textbf{31}(2) (1978), 205--256

\bibitem {ae54}Einstein, A., from a letter to Bohm in Sao Paulo, February 10 1954.

\bibitem {fapa09}Facchi, P. and Pascazio, S.: Quantum Zeno dynamics:
mathematical and physical aspects. \emph{J. Phys.} \textbf{A 41} (2008), 493001--493005

\bibitem {HGoldstein}Goldstein, H.: \emph{Classical Mechanics}.
Addison--Wesley, (1950), 2nd edition, (1980), 3d edition, (2002).

\bibitem {Wiley}de Gosson, M.: \emph{Maslov Classes, Metaplectic
Representation and Lagrangian Quantization}. Research Notes in Mathematics
\textbf{95}, Wiley--VCH, Berlin, 1997.

\bibitem {IHP}de Gosson, M.: On the classical and quantum evolution of
Lagrangian half-forms in phase space. \emph{Ann. Inst. H. Poincar\'{e}},
\textbf{70} (1999), 547--573

\bibitem {principi}de Gosson, M.: \emph{The Principles of Newtonian and
Quantum Mechanics: The need for Planck's constant, $h$.} Imperial College
Press (2001)

\bibitem {Birk}de Gosson, M.: \emph{Symplectic Geometry and Quantum
Mechanics.} Birkh\"{a}user, Basel, series \textquotedblleft Operator Theory:
Advances and Applications\textquotedblright\ (subseries: \textquotedblleft
Advances in Partial Differential Equations\textquotedblright), Vol.
\textbf{166}, (2006)

\bibitem {mdgbh10}de Gosson, M. and Hiley, B. J., Imprints of the Classical
World in Classical Mechanics, Foundations of Physics, \textbf{41}, (2011), 1415-1436.

\bibitem {mdg04PL}de Gosson, M., The optimal pure Gaussian state canonically
associated to a Gaussian quantum state, \emph{Phys. Lett.} \textbf{A330},
(2004), 161-167

\bibitem {gux}Gustafson, K.: A Zeno story. \emph{arXiv:quant-ph/0203032}

\bibitem {Hannabuss}Hannabuss, K.C.: \emph{An introduction to quantum theory}.
Oxford graduate texts in mathematics; 1. Oxford (1997)

\bibitem {wh27}Heisenberg, W, Uber den anschaulichen Inhalt der
quantentheoretischen Kinematik und Mechanik, \emph{Zeit. fur Phsik.},
\textbf{43},(1927) 172-98.

\bibitem {wh49}Heisenberg, W., \emph{The Physical Principles of the Quantum
Theory},pp. 66-76, Dover, New York, (1949)

\bibitem {Hiley1}Hiley, B.J.: Non-Commutative Geometry, the Bohm
Interpretation and the Mind-Mater Relationship, in \emph{Proc. CASYS'2000,}
Li\`{e}ge,\ Belgium, Aug. 7--12, 2000

\bibitem {bh03}Hiley, B.J., Phase Space Descriptions of Quantum Phenomena,
\emph{Proc. Int. Conf. Quantum Theory: Reconsideration of Foundations 2},
267-86, ed. Khrennikov, A., V\"{a}xj\"{o} University Press, V\"{a}xj\"{o},
Sweden, (2003)

\bibitem {bh09}Hiley, B.J., On the Relationship between the Wigner-Moyal and
Bohm Approaches to Quantum Mechanics: A step to a more General Theory.
\emph{Found. Phys.}, \textbf{40} (2010) 365-367. DOI 10.1007/s10701-009-9320-y

\bibitem {hica}Hiley, B.J. and Callaghan, R.E.: Delayed-choice experiments and
the Bohm approach. \emph{Phys. Scr}. \textbf{74} (2006), 336--348

\bibitem {hicama}Hiley, B.J., Callaghan, R.E., Maroney O.J.E.: Quantum
trajectories, real, surreal or an approximation to a deeper process?
\emph{quant-ph/0010020}

\bibitem {Holland}Holland, P.R.:\emph{The quantum theory of motion. An account
of the de Broglie-Bohm causal interpretation of quantum mechanics}. Cambridge
University Press, Cambridge (1995)

\bibitem {Holland1}Holland, P.: Hamiltonian theory of wave and particle in
quantum mechanics I: Liouville's theorem and the interpretation of the de
Broglie-Bohm theory. \emph{Nuovo Cimento} {bf B 116} (2001), 1043--1070

\bibitem {Holland2}Holland, P.: Hamiltonian theory of wave and particle in
quantum mechanics II: Hamilton-Jacobi theory and particle back-reaction.
\emph{Nuovo Cimento} \textbf{B 116}, 1143--1172 (2001)

\bibitem {makmil1}Makri. N., Miller W.H.: Correct short time propagator for
Feynman path integration by power series expansion in $\Delta t$, \emph{Chem.
Phys. Lett}. \textbf{151} (1988), 1-8

\bibitem {makmil2}Makri. N., Miller W.H.: Exponential power series expansion
for the quantum time evolution operator. \emph{J. Chem. Phys.} \textbf{90}
(1989), 904--911

\bibitem {nm29}Mott, N. F., The Wave Mechanics of $\alpha$-Ray Tracks,
\emph{Proc. Roy. Soc.} \textbf{A 126}, (1929), 79-84.

\bibitem {Nelson}Nelson, E.: \emph{Topics in Dynamics I:\ Flows}, Mathematical
Notes, Princeton University Press (1969)

\bibitem {vn55}von Neumann, J., \emph{Mathematical Foundations of Quantum
Mechanics}, Princeton University Press, Princeton (1955).

\bibitem {pdh}Philippidis, C., Dewdney, C. and Hiley, B. J., Quantum
Interference and the Quantum Potential, \emph{Nuovo Cimento}, \textbf{52B},
(1979), 15-28.

\bibitem {primas}Primas, H., \emph{Chemistry, Quantum Mechanics and
Reductionism,} Springer, Berlin (1983)

\bibitem {Schiller1}Schiller, R.: Quasi-Classical Theory of the Nonspinning
Electron. \emph{Phys. Rev.} \textbf{125} (1962), 1100--1108

\bibitem {Schiller2}Schiller, R.: Quasi-Classical Transformation Theory.
\emph{Phys. Rev}. \textbf{125} (1962), 1109--1115

\bibitem {schrint}Schr\"{o}dinger, E.: \emph{The Interpretation of Quantum
Mechanics. Dublin (1949--1955) and other unpublished essays}, edited by Michel
Bitbol. Ox Bow Press, Woodbridge, CT (1995)

\bibitem {Trotter}Trotter, H.F.: On the product of semi-groups of operators.
\emph{Proc. Amer. Math. Soc.} \textbf{10} (1959) 545--551
\end{thebibliography}
\end{document}